\documentclass[letterpaper]{jpconf}

\usepackage{amssymb,amsmath}
\usepackage{graphicx} 
\usepackage{float} 
\usepackage{color}

\def\be{\nopagebreak[3]\begin{equation}}
\def\ee{\end{equation}}
\def\ba{\nopagebreak[3]\begin{eqnarray}}
\def\ea{\end{eqnarray}}

\def\l{\langle}
\def\r{\rangle}
\newcommand{\y}{\hat{y}}              

\newcommand{\py}{{\hat{\pi}^{(y)}}}     
\newcommand{\pyRI}{\hat{\pi}^{(y)R,I}}

\newcommand{\pf}{\hat{\pi}}           
\newcommand{\fluc}[1]{(\Delta #1)^2}      
 
\newcommand{\cre}{\hat{a}^{\dagger}}    

\newcommand{\ann}{\hat{a}}            

\begin{document}
\title{A window to quantum gravity phenomena in the
  emergence of the seeds of cosmic structure}

\author{Daniel Sudarsky$^\dagger$ and Adolfo De
  Unanue$^{\dagger\dagger}$}

\address{Instituto de Ciencias Nucleares, Universidad
  Nacional Aut\'onoma de M\'exico, A. Postal 70-543,
  M\'exico D.F. 04510, M\'exico}

\ead{$^\dagger$sudarsky@nucleares.unam.mx,
  $^{\dagger\dagger}$adolfo@nucleares.unam.mx}

\begin{abstract}
  Inflationary cosmology has, in the last few years,
  received a strong dose of support from observations. The
  fact that the fluctuation spectrum can be extracted from
  the inflationary scenario through an analysis that
  involves quantum field theory in curved space-time, and
  that it coincides with the observational data has lead to
  a certain complacency in the community, which prevents the
  critical analysis of the obscure spots in the derivation.
  We argue here briefly, as we have discussed in more
  detail elsewhere, that there is something important
  missing in our understanding of the origin of the seeds of
  Cosmic Structure, as is evidenced by the fact that in the
  standard accounts the inhomogeneity and anisotropy of our
  universe seems to emerge from an exactly homogeneous and
  isotropic initial state through processes that do not
  break those symmetries.  This article gives a very brief
  recount of the problems faced by the arguments based on
  established physics.  The conclusion is that we need some
  new physics to be able to fully address the problem.  The
  article then exposes one avenue that has been used to
  address the central issue and elaborates on the degree to
  which, the new approach makes different predictions from
  the standard analyses.  The approach is inspired on
  Penrose's proposals that Quantum Gravity might lead to a
  real, dynamical collapse of the wave function, a process
  that we argued has the properties needed to extract us
  from the theoretical impasse described above.
\end{abstract}

\section{Introduction}

One of the big problems facing the search for a unification
of Quantum Theory and Gravitation is the almost complete
absence of experimental data to be used as guidance by
theorists.  In fact there is up to this date no single man
made experiment in which the realms of Quantum Mechanics and
Gravitation intersect, i.e.  are simultaneously needed to
account for the results. The famous COW experiments
\cite{COW} and the recent ``cold neutron" experiments
\cite{GRENOBLE} that are some times exhibited as examples of
tests of the interface of gravity and quantum mechanics, can
not be taken as such when viewed from within the
relativistic paradigm, as they can be fully accounted for in
terms of physics within a single inertial reference frame,
and thus Einsteinian gravity can not be said to be playing
any role (for a more detailed discussion of this point see
for instance \cite{chryssomalakos}).  There exists however {\bf one
  single situation} offered to us by nature, which satisfies
the two criteria of, being observationally accessible and
requiring, for a complete understanding, both general
relativity and quantum physics. That is: the origin of the
seeds of cosmic structure.

Among the most important achievements in observational
cosmology we have the precision measurements of the
anisotropies in the CMB\cite{CMB}.  These together with an
extensive set of observational studies of large scale matter
distribution, led to a very satisfactory picture of the
evolution of structure our universe based on a detailed
understanding of the physics behind it.  In fact it is
nowadays widely accepted that the origin of structure in of
our universe has its natural explanation within the context
of the inflationary scenarios
: Inflation takes relatively arbitrary initial conditions
presumably emerging from a Planck Era and leads to an
\emph{almost} de-Sitter phase of accelerated expansion that
essentially erases all memories of the initial
conditions. At this point inflation has lead to a
featureless universe, which seems to lack even a small
degree of inhomogeneity and anisotropy that is necessary to
lead to the subsequent structure formation.  At this point
quantum mechanics is thought to provide this essential
ingredient: The quantum fluctuations of the inflaton field,
which has been put by inflation in its ``ground
state". These are thought to provide the seeds of the
anisotropies and inhomogeneities that eventually evolve into
the structure we first see in the CMB and eventually into
the obvious features of our universe such as galaxy
clusters, galaxies, stars, etc. The remarkable fact is that
the calculations based on the above scheme seem to lead
naturally to the correct spectrum of these primordial
fluctuations.
  
There is however a serious hole in this seemingly
blemish-less picture: The description of our Universe-- or
the relevant part thereof- starts\footnote{Here we refer to
  the era relevant to the starting point of the analysis
  that leads to the ``fluctuation spectrum".  In the
  standard view of inflation, the relevant region of the
  universe starts with a Planck regime containing large
  fluctuations of essentially all relevant quantities, but
  then, a large number of inflationary e-folds leads to an
  homogeneous and isotropic universe which is in fact the
  starting point of the analysis that takes us to the
  primordial fluctuation spectrum. One might wish, instead,
  to regard such fluctuation spectrum as a remnant of the
  earlier anisotropic and inhomogeneous conditions but then
  one ends up giving up any pretense that one can explain
  its origin and account for its specific form.}  with an
initial condition which is homogeneous and isotropic both in
the background space-time and in the quantum state that is
supposed to describe the ``fluctuations", and it is quite
easy to see that the subsequent evolution through dynamics
that do not break these symmetries can only lead to an
equally homogeneous and anisotropic universe.

In fact many arguments have been put forward in order to
deal with this issue, that is often phrased in terms of the
Quantum to Classical transition --without focusing on the
required concomitant breakdown of homogeneity and isotropy
in the state-- the most popular ones associated with the
notion of decoherence \cite{decoherence}. These the
alternatives have been critically discussed
in\cite{InflationUS, ReplyKP}.  One of the main obstacles is
that in order to justify any explanation based on
decoherence, one has to argue that certain degrees of
freedom must be traced over because they are unobservable,
and this in turn can only be justified by relying on the
limitations we humans currently have, in making certain
measurements.  The problem is that in so doing we would be
using our existence as input, but what cosmology is all
about is understanding the evolution of the universe and its
structure including the emergence of the conditions that
make humans possible.  In other words, in order to
understand the emergence of an inhabitable universe, we
would be relying on the existence and limitations of such
inhabitants.  Therefore any explanation based purely on
standard decoherence becomes circular by definition.  There
are further problems in each of the specific proposals, and
we direct the reader to the above references for extended
discussion of this issues. There are other cosmologists that
have acknowledged that there is problem here, and that
quantum mechanics as we know it needs modifications to be
applicable to the cosmology \cite{Cosmologists2}, with one of
them explicitly stating that decoherence does not offer a
complete and satisfactory resolution to this
problem \cite{Mukhanov}.

Moreover, if we were to think in terms of first principles,
we would start by acknowledging that the correct description
of the problem at hand would involve a full theory of
quantum gravity coupled to a theory of all the matter
quantum fields, and that there, the issue would be whether
we start with a quantum state that is homogeneous and
isotropic or not?. Even if these notions do not make sense
within that level of description, a fair question is whether
or not, the inhomogeneities and anisotropies we are
interested on, can be traced to aspects of the description
that have no contra-part in the approximation we are using.
Recall that such description involves the separation of
background \emph{vs.} fluctuations and thus must be viewed only as
an approximation, that allows us to separate the
nonlinearities in the system--as well as those aspects that
are inherent to quantum gravity-- from the linear part of
problem represented by the fluctuations, which are treated
in terms of linear quantum field theory.  In this sense, we
might be tempted to ignore the problem and view it as
something inherent to such approximation.  This would be
fine, but we should recognize then that we could not argue
that we understand the origin of the CMB spectrum, if we
view the asymmetries it embodies as arising from some aspect
of the theory we do not know, rely on, or touch upon.  In
fact, in the treatment which we describe next, the proposal
is to bring up one particular element or aspect, that we
view as part of the quantum gravity realm, to the forefront
of the treatment, in order to modify --in a minimalistic
way-- the semiclassical treatment, that, as we said, we find
lacking, and provide a setting in which the obscure issues
can at least be focus on.  It is of course not at
all clear that the problem we are discussing should be
related to quantum gravity, but the later is the only sphere
which is now believed capable of leading to a radical change
in the paradigm of fundamental physics which is of course
what we are considering here.

The approach taken in \cite{InflationUS} is influenced by
Penrose's suggestion that quantum gravity might play a role
in triggering a real dynamical collapse of the wave function
of systems \cite{Penrose}. His proposals would have a system
collapsing whenever the gravitational interaction energy
between two alternative realizations that appear as
superposed in a wave function of a system reaches a
threshold naturally identified with $M_{Planck}$.  We will
show that these ideas can, in principle, be investigated in
the present context , and that they could lead to observable
effects. In fact the very early universe
can be seen as a case for which there exists already a
wealth of empirical information and one which, as we have
argued can not be fully understood without involving some
New Physics, with features that would seem to be quite close
to those of Penrose's proposals.  There are of course
alternatives settings in which modifications of quantum
theory, in principle unrelated to Quantum Gravity, could
play the role of the New Physics\cite{GRW} that we have
argued is needed in order to account for the seeds of cosmic
structure, but we will limit ourselves here to a rather
generic setting motivated by the first set of ideas.

\section{ Quantum Gravity and its effective semiclassical
  description}\label{sec_qg}
Before we present the treatment we are proposing, which
should be consider as being of phenomenological nature it is
worthwhile to see how would it fit within the context of a
fundamental theory such as a quantum theory of gravity.
  
The first thing one should note is that the notions of
space-time are likely to change dramatically when considered
in a fully quantum theory of gravitation\cite{Notions}.  A
fundamental theory of quantum gravity (with or without
matter) is naturally expected to be a timeless
theory\footnote{There are of course approaches that do
  otherwise like String Theory (see for instance \cite{ST})
  or the Causal Sets program \cite{POSET}.},
but General Relativity is certainly not.

In fact we have a good example of this arising in the past
and current attempts to apply the canonical quantization
procedure to the theory of General Relativity: In all such
schemes one ends with a timeless theory in which the wave
functionals $\Psi [h_{ab}] $ depend on (see
\cite{Canonical}) a spatial metric $h_{ab}$ associated with
a slice $\Sigma $ of space-time (its canonical conjugate
momentum $\pi^{ab}$ is a closely related to the extrinsic
curvature of such slice $K_{ab}$), or, equivalent variables
such as triads and connections as in the modern incarnation
of the program known as Loop Quantum Gravity (see for
example \cite{LQG}).
  
The recovery of the usual space-time notions --that in the
appropriate limit should be GR-- is thus expected to proceed
trough a rather indirect way which involves the
identification of one of the variables of the formalism with
something playing the role of a physical clock\cite{Gambini
  Pullin}, and in fact this seems to be more easily achieved
in a theory involving not only the gravitational degrees of
freedom, but also some suitable matter fields \cite{Fields
  as Clocks}, for instance a scalar field $\phi$, where one
would have a wave functional $\Psi[ h_{ab}, \phi]$.  The
idea is then that upon the identification of $\phi$ as a
physical clock variable, one would be able to talk about the
probability that the space-time metric and its conjugate
variable take such and such value when the clock takes a
given value, and from such information one would presumably
be able to estimate the most likely space-time, correlations
and so forth.  Examples of application of such approach can
be seen in \cite{Relational Time}.  These effective
descriptions can be expected quite generally to incorporate
some degree of breakdown in unitary evolution
\cite{Gambini Pullin}.
 
Our point here is that the recovering of the usual
space-time picture is expected, to be a complex procedure
even if we have a complete theory of quantum gravity in
interaction with all matter fields.  It should be said that
in the LQG program, even the simpler spatial notions as
distance volumes and to a lesser degree areas, turn out to
be also a rather involved procedure.
  
The recovery of the standard evolution of physical degrees
of freedom in space-time can be expected to involve an even
more cumbersome procedure including suitable approximations
and averaging.  It is thus not unnatural to consider that
those might include what we would call ``jumps", and in
general the sort of general behavior that would look as an
effective collapse of ``the wave-function" as seen from the
stand point of the effective description. Needless is to say
that we have at this time no hope of being able to describe
the above procedure in any detail, among other reasons
because we do not have at the moment a fully satisfactory
and workable theory of quantum gravity.
 
On the other hand the effective description is expected to
lead in the appropriate limit to General Relativity as the
description of space-time, and in the corresponding
appropriate limit, to quantum field theory for the
description of matter fields. We assume that the situation
of interest (the inflationary regime of the early universe)
lies in a region where both these descriptions are
approximately valid, but where some modifications tied to
the fact that these picture is only an effective one, need
to be incorporated in a seemingly {\it ad hoc} manner. Of
course we would be complete loss if we did not have any
other guidance as to the nature of these modifications, but
here is where we look in the opposite direction and guide
ourselves on some empirical facts: The inflationary account
of the origin of the seeds of cosmic structure works "almost
well" and its defects might be dealt with by the
introduction of one such additional feature to the picture.
What is required is the assumption of the existence of a
process that can take a symmetric (i.e H\&I) state of a
close system (the Universe), into a state with small
departures from a symmetric state while the standard
evolution would have preserved those symmetries.  As we will
see, a quantum mechanical collapse of a wave function seems
to have the required characteristics, except that it is
usually assumed to be associated with the interaction of the
quantum mechanical system with an external classical
``apparatus" or ``observer".  It is clear that in the
situation at hand we can not call upon any such feature so
we will assume that the feature in question appears as a
self induced collapse, along the lines that have been
suggested, based on quite different arguments, as a likely
feature of quantum gravity by R. Penrose.

These observations and ideas lead us to consider, situations
where a quantum treatment of other fields would be
appropriate but an effective classical treatment of
gravitation would be justified.  That is the realm of
semi-classical gravity that we will assume to be valid in
our context except at those instants where would break down
in association with the jumps or collapses of the state of
the quantum field that we considered to be part of the
effective description of underlying fundamental quantum
theory containing gravitation.

In accordance with the ideas above we will use a
semi-classical description of gravitation in interaction
with quantum fields as reflected in the semi-classical
Einstein's equation $R_{ab} -(1/2) g_{ab} R = 8\pi G \l
T_{ab} \r $ whereas the other fields are treated in the
standard quantum field theory (in curved space-time)
fashion. As indicated this could not hold when a quantum
gravity induced collapse of the wave function occurs, at
which time, the excitation of the fundamental quantum
gravitational degrees of freedom must be taken into account,
with the corresponding breakdown of the semi-classical
approximation. The possible breakdown of the semi-classical
approximation is formally represented by the presence of a
term $Q_{ab} $ in the semi-classical Einstein's equation
which is supposed to become nonzero {\bf only} during the
collapse of the quantum mechanical wave function of the
matter fields. Thus we write \be
\label{SemiCEQ}
R_{ab} -(1/2) g_{ab} R +Q_{ab} =8\pi G \l T_{ab} \r \ee
Thus, we consider the development of the state of the
universe during the time at which the seeds of structure
emerge to be initially described by a H.\& I. state for the
gravitational and matter D.O.F.  At some time, $\eta_c$, the
quantum state of the matter fields reaches a stage whereby
the corresponding state for the gravitational D.O.F. is
forbidden, and a quantum collapse of the matter field wave
function is triggered.  This new state of the matter fields
does no longer need to share the symmetries of the initial
state, and by its connection to the gravitational D.O.F. now
accurately described by Einstein's semi-classical equation
leads to a geometry that is no longer homogeneous and
isotropic.

That is the approach that is taken in the work presented
here, where the intent will be to emphasize the
phenomenology potential and the lessons that can be
extracted from it in order to make it clear that the
proposal can be viewed as a viable path to investigate
aspects of the physical world that have for long seemed
absolutely beyond reach.

\section{ The quantum origin of the seeds of cosmic
  structure}\label{sec_main}


Next we give a short description of the analysis of the
origin of the primordial cosmological inhomogeneities and
anisotropies based on the ideas outlined above. The staring
point is as usual the action of a scalar field coupled to
gravity.  \be
\label{eq_action}
S=\int d^4x \sqrt{-g} \lbrack \frac{1}{16\pi G} R[g] -
1/2\nabla_a\phi \nabla_b\phi g^{ab} - V(\phi)\rbrack, \ee
where $\phi$ stands for the inflaton and for its potential
$V$.  One then splits both, metric and scalar field into a
spatially homogeneous ``background'' part and an
inhomogeneous part ``fluctuation'', i.e. the scalar field is
written $\phi=\phi_0+\delta\phi$, while the perturbed metric
can, (after appropriate gauge fixing and by focusing on the
scalar perturbation) be written
\begin{equation}
  ds^2=a(\eta)^2\left[-(1+ 2 \Psi) d\eta^2 + (1- 2
    \Psi)\delta_{ij} dx^idx^j\right],
\end{equation}
where $\Psi$ is the relevant perturbation called the
"Newtonian potential".

The background solution corresponds to the standard
inflationary cosmology during the inflationary era has a
scale factor $a(\eta)=-\frac{1}{H_{\rm I} \eta},$ with $ H_I
^2\approx (8\pi/3) G V$while the scalar $\phi_0$ field in
the slow roll regime so $\dot\phi_0= - (a^3/3 \dot a)V'$.
 
The perturbation of the scalar field leads to a perturbation
of the energy momentum tensor, and thus Einstein's equations
at lowest order lead to
\begin{equation}
  \nabla^2 \Psi  = 4\pi G \dot \phi_0 \delta\dot\phi  \equiv s \delta\dot\phi
  \label{main3}
\end{equation}
where $s=4\pi G \dot \phi_0$.

We must now write the quantum theory of the rescaled the
field $y=a \delta \phi$.  For definiteness we consider the
field in a box of side $L$, and write the field and momentum
operators as
\begin{equation}
  \y(\eta,\vec{x})=
  \frac{1}{L^{3}}\sum_{\vec k}\ e^{i\vec{k}\cdot\vec{x}} \hat y_k
  (\eta), \qquad \py(\eta,\vec{x}) =
  \frac{1}{L^{3}}\sum_{\vec k}\ e^{i\vec{k}\cdot\vec{x}} \hat \pi_k
  (\eta),
\end{equation}
where the sum is over the wave vectors $\vec k$ satisfying
$k_i = 2\pi n_i /L$ for $i=1,2,3$ with $n_i$ integers, then,
we write the fields in terms of the annihilation and
creation operators $\hat y_k (\eta) \equiv y_k(\eta) \ann_k
+\bar y_k(\eta) \cre_{-k}$ and $\hat \pi_k (\eta) \equiv
g_k(\eta) \ann_k + \bar g_{k}(\eta) \cre_{-k}$ with
\begin{equation}
  y^{(\pm)}_k(\eta)=\frac{1}{\sqrt{2k}}\left(1\pm\frac{i}{\eta
      k}\right)\exp(\pm i k\eta),\qquad
  g^{\pm}_k(\eta)=\pm
  i\sqrt{\frac{k}{2}}\exp(\pm i k\eta) . \label{Sol-g} 
\end{equation}
Given that we are interested in considering a kind of self
induced collapse which operates in close analogy with a
``measurement" which normally involves self adjoint
operators, we work find it convenient with the real and
imaginary components of the fields and thus we write $\hat
y_k (\eta)=\hat y_k{}^R (\eta) +i \hat y_k{}^I (\eta)$ and
$\hat \pi_k (\eta) =\hat \pi_k{}^R (\eta) +i \hat \pi_k{}^I
(\eta)$ where the operators $\hat y_k^{R, I} (\eta)$ and
$\hat \pi_k^{R, I} (\eta)$ are hermitian. Let $|\Xi\rangle$
be any state in the Fock space of $\hat{y}$,and assign to
each such state the following quantity: $d_k^{R,I}= \l
\ann_k^{R,I} \r_\Xi.$ The expectation values of the modes of
the fundamental field operators are then expressible as
\begin{equation}\label{eq:expectation}
  \l {\y_k{}^{R,I}} \r_\Xi = \sqrt{2} \Re (y_k d_k^{R,I}),  \qquad
  \l {\py_k{}^{R,I}} \r_\Xi = \sqrt{2} \Re (g_k d_k^{R,I}).
\end{equation}

For the vacuum state $|0\rangle$ we have of course: $
\l{\y_k{}^{R,I}}\r_0 = 0, \l\py_k{}^{R,I}\r_0 =0, $ while
their corresponding uncertainties are
\begin{equation}\label{momentito}
  \fluc{\y_k {}^{R,I}}_0 =(1/2) |{y_k}|^2(\hbar L^3), \qquad
  \fluc{\pf_k {}^{R,I}}_0 =(1/2)|{g_k}|^2(\hbar L^3).
\end{equation}
 
{\bf The collapse:}
Next we provide a simple specification of what we mean by
``the collapse of the wave function" by stating the form
collapsed state in terms of its collapse time.  We assume
the collapse to be analogous to some sort of imprecise
measurement of the operators $\hat y_k^{R, I} (\eta)$ and
$\hat \pi_k^{R, I} (\eta)$. In order to describe is the
state $|\Theta\rangle$ after the collapse we must specify
$d^{R,I}_{k} = \langle\Theta|\ann_k^{R,I}|\Theta\rangle $.
This is done by making the following assumption about the
state $|\Theta\rangle$ after collapse:
\begin{subequations}\label{Schemme1}
\be
  \l {\y_k^{R,I}(\eta^c_k)} \r_\Theta=x^{R,I}_{k,1}
  \sqrt{\fluc{\y^{R,I}_k}_0}=x^{R,I}_{k,1}|y_k(\eta^c_k)|\sqrt{\hbar
    L^3/2},
\ee
\be
  \l {\py_k{}^{R,I}(\eta^c_k)}\r_\Theta=x^{R,I}_{k,2}\sqrt{\fluc{\pyRI_k}
    _0}=x^{R,I}_{k,2}|g_k(\eta^c_k)|\sqrt{\hbar L^3/2},
\ee
\end{subequations}
where $x_{k,1},x_{k,2}$ are selected randomly from within a
Gaussian distribution centered at zero with spread one. We
note that our universe, corresponds to a single realization
of the random variables, and thus each of the quantities $
x^{R,I}{}_{k,1,2}$ has a single specific value.  Later, we
will see how to make relatively specific predictions,
despite these features.

The connection to gravitational sector is at the
semi-classical level so Eq.(\ref{main3}) turns into
\begin{equation}
  \nabla^2 \Psi = s \langle \delta\dot\phi\rangle. \label{main4}
\end{equation}
We note that before the collapse, the expectation value on
the right hand side is zero. Next we determine what happens
after the collapse: To this end, we need to solve the
equations \eqref{Schemme1} for $d_k$ and then substitute
this in \eqref{eq:expectation} and then in the
Fourier transform of Eq.(\ref{main4}) and obtain
\begin{equation}\label{Psi}
  \Psi_k(\eta)=\frac{-s}{k^2}\langle \delta\dot\phi_k\rangle_\Theta=\frac{-s}{k^2}\sqrt{\hbar L^3 k}\frac{1}{2a}F(k), 
\end{equation}
where
\begin{equation}\label{F}
  F(k) = (1/2) [A_k (x^{R}_{k,1} +ix^{I}_{k,1}) + B_k (x^{R}_{k,2}
  +ix^{I}_{k,2})],
\end{equation}
with
\begin{equation} A_k = \frac {\sqrt{ 1+z_k^2}} {z_k}
  \sin(\Delta_k) , \qquad B_k =\cos (\Delta_k) + (1/z_k)
  \sin(\Delta_k),
\end{equation}
and where $\Delta_k= k \eta -z_k$ with $ z_k =\eta_k^c k$.

\begin{figure*}
  \includegraphics[width=150mm,height=130mm]{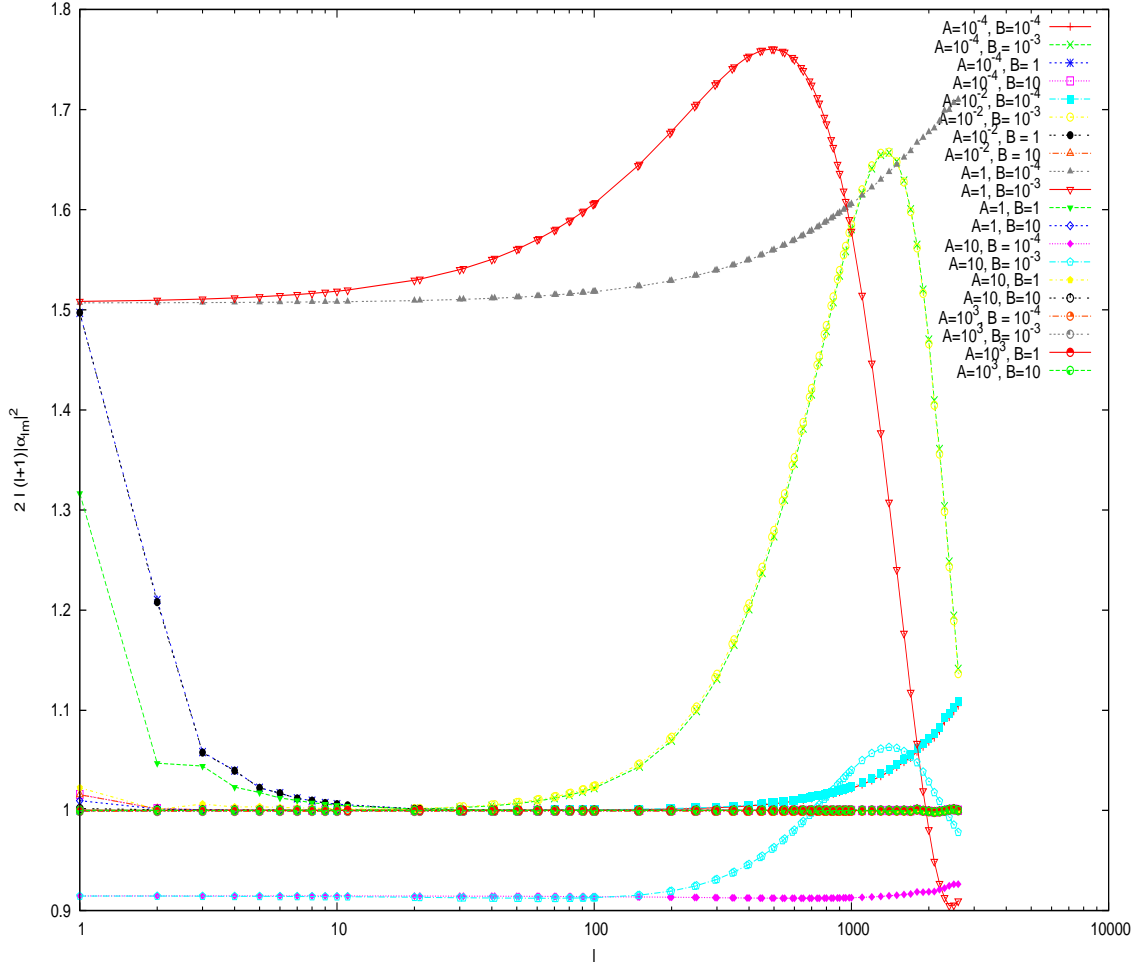}
  \caption{Semi-log plot of $|\alpha_{lm}|^2(C_1(k))$
    assuming a small departure from the result ``$z_k$
    independent of $k$''. This small departure is
    represented by $\tilde z_x = A + Bx$, where $x = kR_D$,
    with $R_D$ standing for the radius of last scattering
    surface.  Here, we are showing different values of
    $(A,B)$, representing how robust is the scheme of
    collapse under this circumstances. The abscissa ranges from $l=0$
    to $l=2600$. See section \ref{sec:furth-phen-analys}
    for more details about the meaning of the variables $A$
    and $B$.}
  \label{fig:c1_log}
\end{figure*}

Turning to the observational quantities we recall that the
quantity that is measured is $\frac{\Delta T}{T}$ as a
function of $(\theta,\varphi)$, the coordinates on the
celestial two-sphere which is expressed as $\sum_{lm}
\alpha_{lm} Y_{l,m}(\theta,\varphi)$.  The angular
variations of the temperature are then identified with the
corresponding variations in the ``Newtonian Potential" $
\Psi$, by the understanding that they are the result of
gravitational red-shift in the CMB photon frequency $\nu$ so
$\frac{\delta T}{T}=\frac{\delta \nu}{\nu} = \frac{\delta
  (\sqrt{g_{00}})}{\sqrt{g_{00}}} \approx \Psi$ (we are
ignoring, for simplicity the complications of the late time
physics such as reheating or acoustic oscillations).  Thus,
the quantity of interest is the ``Newtonian potential" on
the surface of last scattering: $\Psi(\eta_D,\vec{x}_D)$,
from where one extracts $a_{lm}=\int \Psi(\eta_D,\vec{x}_D)
Y_{lm}^* d^2\Omega.$ To evaluate the expected value for the
quantity of interest we use (\ref{Psi}) and (\ref{F}) to
write
\begin{equation}
  \Psi(\eta,\vec{x})=\sum_{\vec k}\frac{s  } {k^2}\sqrt{\frac{\hbar
      k}{L^3}}\frac{1}{2a}
  F(\vec{k})e^{i\vec{k}\cdot\vec{x}},
  \label{Psi2}
\end{equation}
then, after some algebra we obtain
\begin{eqnarray}
  \alpha_{lm}&=&s\sqrt{\frac{\hbar}{L^3}}\frac{1}{2a} \sum_{\vec
    k}\frac{U(k)\sqrt{k}}{k^2} F(\vec k)  4 \pi i^l  j_l((|\vec k|
  R_D) Y_{lm}(\hat k),\label{alm1}
\end{eqnarray}
where $\hat k$ indicates the direction of the vector $\vec
k$. It is in this expression that the justification for the
use of statistics becomes clear.  The quantity we want to
evaluate is the result of the combined contributions of an
ensemble of collapsing harmonic oscillators each one
contributing with a complex number to the sum, leading to
what is in effect a bi-dimensional random walk whose total
displacement corresponds to the observational quantity. We
can not of curse evaluate such total displacement but only
its most likely value We do so and then take the continuum
limit and which after rescaling the variable of integration
to $x =kR_D$, becomes
\begin{equation} \label{alm5}
  |\alpha_{lm}|^2_{M. L.}=\frac{s^2 \hbar}{2 \pi a^2} \int
  \frac{C(x/R_D)}{x^4} j^2_l(x) x^3 dx,
\end{equation}
where
\begin{equation}
  C(k)\equiv 1+ (2/ z_k^2) \sin (\Delta_k)^2 + (1/z_k)\sin (2\Delta_k).
  \label{ExpCk}
\end{equation}
In the exponential expansion regime where $\mu$ vanishes and
in the limit $z_k\to -\infty$ where $C=1$, we find:
\begin{equation}
  |\alpha_{lm}|^2_{M. L.}=\frac{s^2    \hbar} {2  a^2}
  \frac{1}{l(l+1)} .
\end{equation}
which has the standard functional result.  However we must
consider the effect of the finite value of times of collapse
$\eta^c_k$ codified in the function $C(k)$. We note is that
in order to get a reasonable spectrum there is a single
simple option: That $z_k $ be essentially independent of $k$
that is the time of collapse of the different modes should
depend on the mode's frequency according to $\eta_k^c=z/k$.
There are of course other possible schemes of collapse and
we have investigated the most natural ones and their
corresponding effects on the primordial fluctuation
spectrum, with results that to a large extent confirm that
the above conclusion is rather robust (see figure
\ref{fig:c1_log}, section \ref{sec:furth-phen-analys} and
specifically \cite{Adolfo} for a deeper discussion). Thus we
can conclude that the above pattern of times of collapse
seems to be implied by the data (as far as our preliminary
analysis has shown so far). In our view such conclusion
represents one important and relevant piece of information
about whatever the mechanism of collapse is.

\section{ A version of `Penrose's mechanism' for collapse in
  the cosmological setting}
\label{sec_penrose}
Based on the analysis of the inadequacies of Quantum
Mechanics as a complete theory of nature and the places from
where solutions can arise, R.  Penrose has argued that the
collapse of quantum mechanical wave functions is an actual
dynamical process, independent of observation, and that the
fundamental physics is related to quantum gravity. More
precisely, according to this suggestion, the collapse into
one of several coexisting quantum mechanical alternatives
would take place when the gravitational interaction energy
between the alternatives exceeds a certain threshold.  We
have considered a naive realization of Penrose's ideas
appropriate for the present setting to be as follows: Each
mode would collapse by the action of the gravitational
interaction between it's own possible realizations. In our
case, one could estimate the interaction energy
$E_I(k,\eta)$ by considering two representatives of the
possible collapsed states on opposite sides of the Gaussian
associated with the vacuum. We interpret $\Psi$, literally
as the Newtonian potential and consequently , $\rho=
a^{-2}\dot\phi_0 \delta\dot\phi $ should be identified with
matter density.  Then the gravitational interaction energy
between alternatives should be: \be\label{GE1}
E_I(\eta)=\int \Psi^{(1)} \rho^{(2)}dV = (a/L^3)\dot\phi_0
\Sigma_{k} \Psi^{(1)}_{ k}(\eta)
\delta\dot\phi^{(2)}_{k}(\eta) , \ee where $(1),(2)$ refer
to the two different realizations chosen. Recalling that
$\Psi_{ k} = (-s/k^2) \delta\dot\phi_k$, we find \be
E_I(\eta)= -4\pi G (a/L^3) \dot\phi_0^2\Sigma_{k} (1/k^2)
\delta\dot\phi^{(1)}_{k}(\eta)
\delta\dot\phi^{(2)}_{k}(\eta)\approx \Sigma_{k}(\pi \hbar
G/ak) (\dot\phi_0)^2.  \ee Where we have used equation
(\ref{momentito}), to estimate $
\delta\dot\phi^{(1)}_{k}(\eta)
\delta\dot\phi_{k}^{(2)}(\eta) $ by $|< \delta\dot\phi_k >
|^2 = \hbar k L^3 (1/2a)^2$.

This result can be interpreted as the sum of the
contributions of each mode to the interaction energy of
different alternatives.  We view each mode's collapse as
occurring independently although at this point it is rather
unclear if this can be fully justified, and thus, the
collapse of mode $k $ would occur when this energy
$E_I(k,\eta)=(\pi \hbar G/ak) (\dot\phi_0)^2 =\frac{ \pi
  \hbar G } { 9H_I^2} (a/k) (V')^2$ reaches the value of the
Planck Mass $M_p$.  Thus the condition determining the time
of collapse $\eta^c_k$ of the mode $k$ becomes, \be
z_k=\eta^c_k k =\frac{\pi }{9} (\hbar V')^2(H_I
M_p)^{-3}=\frac{\epsilon} {8\sqrt {6\pi}}(\tilde
V)^{1/2}\equiv z^c, \ee which is independent of $k$, and
thus, as we saw in the previous section leads to a roughly
scale invariant spectrum of fluctuations in accordance with
observations.

\section{Further phenomenological
  analysis}\label{sec:furth-phen-analys}
The scheme of collapse we have considered in section
\ref{sec_main}, and which will be referred as ``the
symmetric scheme" or scheme No 1, is evidently far from
unique and other similarly natural schemes can be considered
here we will briefly discuss two alternatives: ``the
momentum preferred scheme" or scheme No 2 and the ``Wigner
functional scheme" or scheme No 3.  The first corresponds to
the assumption that it is only the momentum conjugate mean
value that changes during the collapse according to equation
\ref {Schemme1} while the field's expectation value
maintains its initial value during the collapse, namely
zero.  The scheme No 3 corresponds to the assumption that
after the collapse the expectation values of field and
momentum modes, follow the correlations in the corresponding
uncertainties that existed in the of the pre-collapse state,
namely:
\begin{equation}
  \label{eq:collapse_scheme_wigner}
  \left\langle\hat{y}_k^{(R,I)}\left(\eta_k^c\right)\right\rangle_\Omega
  = x^{(R,I)}_k \Lambda_k
  \cos\Theta_k\, ,
  \quad
  \left\langle\hat{\pi}_k^{(R,I)}\left(\eta_k^c\right)\right\rangle_\Omega
  = x^{(R,I)}_k \Lambda_k k
  \sin\Theta_k,
\end{equation}
where $\Lambda_k$ is given by the major semi-axis of the
ellipse characterizing the bi dimensional Gaussian function
(the ellipse corresponds to the boundary of the region in
``phase space'' where the Wigner function has a magnitude
larger than $ 1/2$ its maximum value), and $\Theta_k $ is
the angle between that axis and the $y_k^{R,I }$ axis.

The subsequent analysis proceeds in the same fashion as that
presented in section \ref{sec_main} and the result has the
same mathematical expression as in equation (\ref{alm5}),
with the sole exception being the exact expression of the
function $C(k)$, which for the scheme no. 2 takes the form:
\begin{equation}\label{C_2}
  C_2(k) = 1 + \sin^2\Delta_k \left(1 - \frac{1}{z_k^2}\right)
  - \frac{1}{z_k}\sin (2\Delta_k),
\end{equation}
and in the scheme no. 3 is:
\begin{multline}
  \label{C_wigner_2}
  C_{3}(k) = \frac{32 z_k^2}{\mathcal{A}
    (1+5z_k^2-\mathcal{A})} \times \\
  \Bigg\{ \left[\mathcal{A} - 1
    +3z_k^2\right]\left(\cos\Delta_k
    -\frac{\sin\Delta_k}{z_k}\right)^2 + \sin^2\Delta_k
  \left[\mathcal{A} - 3z_k^2 - 7 \right] + 8z_k\cos\Delta_k
  \sin\Delta_k \Bigg\}.
\end{multline}
with $\mathcal{A} = \sqrt{1 + 10z_k^2 + 9z_k^4}$. In
\cite{Adolfo} it was shown that, despite the fact that the
expression for $C_{3}$ looks by far more complicated that
$C_2$, their dependence in $z_k$ is very similar, except for
the amplitude of the oscillations.
 
These in turn lead to particular forms of the primordial
spectrum (i.e the spectrum which emerges from inflation and
has not yet been modified to include the late time physics
such as the acoustic oscillations responsible for the famous
peaks).  At the approximation level we are working here the
spectra would be all identical to the standard ($n=1$) scale
invariant Harrison-Zel'dovich (HZ) spectrum corresponding to
a flat line in the graph of $2l(l+1)|\alpha_{lm}|^2 $
vs. $l$, if we assume that $z_k$ is independent of $k$
(corresponding to a time of collapse of mode $k$ given by $
\eta_k^c = z/k$). Therefore, at this level the different
collapse schemes are not distinguishable.
 
We thus were let to considered the sensitivity of the
resulting spectrum to small deviations of the ``$z_k$
independent of $k$ pattern'' by studying a linear
departure from the $k$ independent $z_k$ characterized by
$\tilde{z}_x$ as $\tilde{z}_x = A + Bx$ ($x= k R_D$ where
$R_D$ stands for the radius of the surface of s last
scattering) in order to examine the robustness of the
various collapse schemes in as far as predicting the
observational spectrum.

\begin{figure*}
  \includegraphics[width=150mmm, height=130mm]{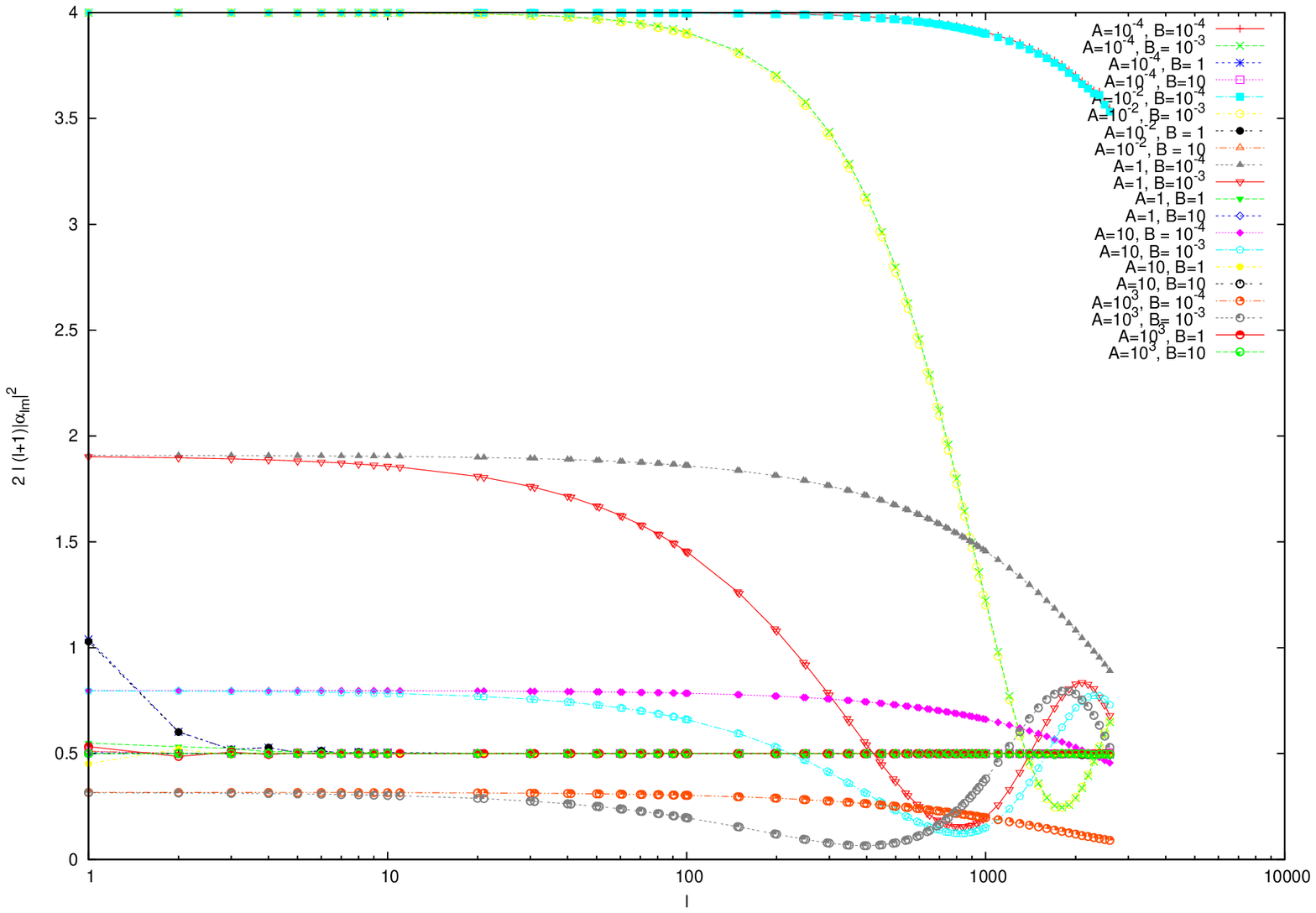}
  \caption{Semi-log plot of $|\alpha_{lm}|^2(C_2(k))$ for
    different values of $(A,B)$, representing how robust is
    the scheme of collapse when it departs from $z_k$
    constant.The abscissa ranges from $l = 0$ to $l=2600$}
  \label{fig:c2_log}
\end{figure*}

\begin{figure*}
  \includegraphics[width=150mm,
  height=130mm]{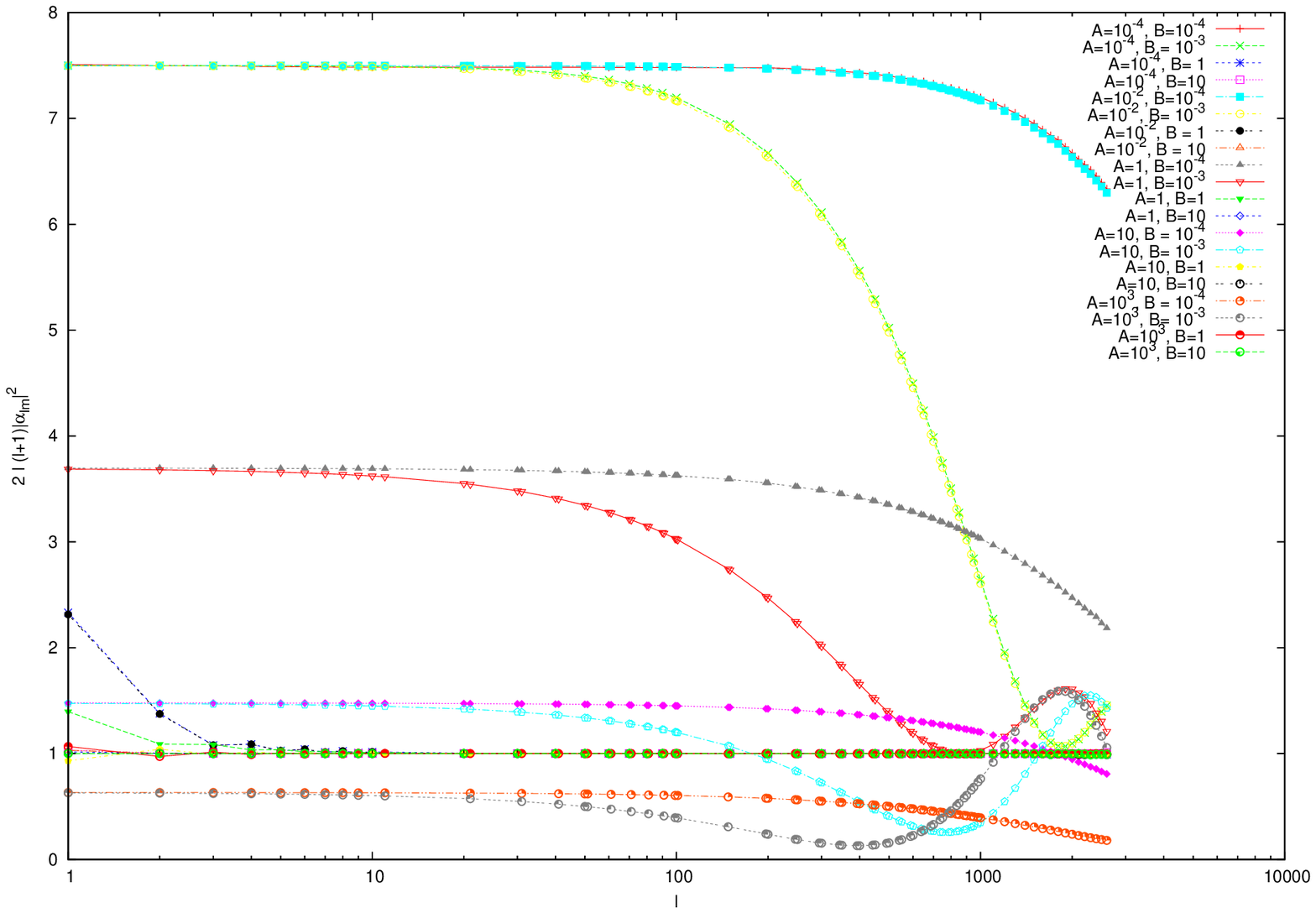}
  \caption{Semi-log plot of $|\alpha_{lm}|^2(C_3(k))$ for
    different values of $(A,B)$, representing how robust is
    the scheme of collapse when it departs from $z_k$
    constant. The abscissa is $l$ up to $l=2600$.}
  \label{fig:c_wigner_log}
\end{figure*}

Some of the results of these analysis can be seen in the
graphs \ref{fig:c2_log}, \ref{fig:c_wigner_log} in which we
see that each collapse scheme leads to a particular pattern
of modifications of the spectrum, clearly showing the
potential of the present approach to teach us something
about the effective collapse, or alternatively to account
for possible deviations if anything of this sort were to be
detected in future observations.

One of the most important predictions of the scheme, is the
absence of tensor modes, or at least their very strong
suppression.  This can be understood by considering the
semi-classical version of Einstein's equation and its role
in describing the manner in which the inhomogeneities and
anisotropies in the metric arise in our scheme.  As
indicated in the introduction, the metric is taken to be an
effective description of the gravitational D.O.F., in the
classical regime, and not as the fundamental
D.O.F. susceptible to be described at the quantum level.  It
is thus the matter degrees of freedom (which in the present
context are represented by the inflaton field) the ones that
are described quantum mechanically and which, as a result of
an hypothetical fundamental aspect of gravitation at the
quantum level, would be subject to an effective quantum
collapse (the reader should recall that our point of view is
that gravitation at the quantum level will be drastically
different from standard quantum theories, and that, in
particular, it will not involve universal unitary
evolution). This leads to a nontrivial value for $\l T_{ab}
\r$, which leads to the appearance of the metric
fluctuations.  The point is that the energy momentum tensor
contains linear and quadratic terms in the expectation
values of the quantum matter field fluctuations, which are
the source terms determining the geometric perturbations. In
the case of the scalar perturbations, there are first order
contributions to the perturbed energy momentum tensor, which
are proportional to $\dot\phi_0\l \delta\dot\phi \r$, while
there are no similar first order terms that would appear as
source of the tensor perturbations (i.e. of the
gravitational waves).  In the usual treatment, and besides
its conceptual shortcomings, no such natural suppression of
the tensor modes can be envisaged. At the time of the
writing of this article, the tensor modes had not been
detected, in contrast with the scalar modes.

\section{Discussion}
      
We have presented the first steps in the proposal involving
the introduction of novel aspect (a self induce collapse of
the wave function) of physics in the description of the
emergence from the quantum uncertainties in the state of the
inflaton field, of the seeds of structure in our
universe. We have argued that such novel aspect is likely to
be associated to the connection of a fundamental theory of
quantum gravity, with the effective description in terms of
the equations of semi-classical general relativity. We find
it quite remarkable that in doing so, we are able to obtain
a relatively satisfactory picture.  We do not know what
exactly is the physics of collapse but we were nevertheless
able to obtain some constraints on it (about the time of
collapse of the different modes), and shown that a
simplistic extrapolation of Penrose's ideas satisfy this
constraint. We have not investigated the possible connection
of our proposal with other more developed schemes involving
similar non unitary modifications of quantum theory such as
the various schemes considered by colleagues participating
in this meeting and others.  Te reason for that is that we
found it better to try to extract some information about
what would be needed for the scheme to work in the
cosmological case on which we have centered our interest,
and would hope to be able to explore the connections of our
proposal, with such schemes an their compatibility of their
implications with the conclusions extracted in this initial
analysis.
 
We have reviewed the serious shortcoming of the inflationary
account of the origin of cosmic structure, and have given a
brief account of the proposals to deal with them which were
first reported in \cite{InflationUS}.  These lines of
inquiry have lead to the recognition that something else
seems to be needed for the whole picture to work and that it
could be pointing towards an actual manifestation quantum
gravity.  We have shown that not only the issues are
susceptible of scientific investigation based on
observations, but also that a simple account of what is
needed, seems to be provided by the extrapolation of
Penrose's ideas to the cosmological setting.  Interestingly
the scheme does in fact lead to some deviations from the
standard picture where the metric and scalar field
perturbations are quantized.  For instance, as discussed in
the last section, one is lead to expect no excitation of the
tensor modes, something that we can expected to be able
confront with relatively precise data in the near future.

We also find new avenues to address the fine tuning problem
that affects most inflationary models, because one can
follow in more detail the objects that give rise to the
anisotropies and inhomogeneities, and by having the
possibility to consider independently the issues relative to
formation of the perturbation, and their evolution through
the reheating era (for a more extended discussion of this
point see \cite{Napflio}).

Other aspects that can, in principle, be tested, were
discussed in the last section.  The noteworthy fact is that
what initially could have been thought to be essentially a
philosophical problem, leads instead to truly physical
issues.
   
Our main point is however that in our search for physical
manifestations of new physics tied to quantum aspects of
gravitation, we might have been ignoring what could be the
most dramatic such occurrence: The cosmic structure of the
Universe itself.


\ack It is a pleasure to acknowledge very helpful
conversations with J. Garriga, E. Verdaguer and
A. Perez. This work was supported in part by DGAPA-UNAM
IN108103 grant.

\end{document}